# Imaging Electron Interferometer

B.J. LeRoy[1,*], A.C. Bleszynski[1], K.E. Aidala[2], R.M. Westervelt[1,2], A. Kalben[3], E. J. Heller[1,3],

S. E. J. Shaw[4], K.D. Maranowski[5] and A.C. Gossard[5]

[1]Department of Physics, [2]Division of Engineering and Applied Sciences and

[3]Department of Chemistry and Chemical Biology, Harvard University, Cambridge, MA 02138

[4]MIT Lincoln Laboratory, Lexington, MA 02420

[5]Materials Department, University of California, Santa Barbara, CA 93106

[*]Current address: Kavli Institute of Nanoscience, Delft University of Technology,

2628 CJ Delft, The Netherlands

## Abstract

An imaging interferometer was created in a two-dimensional electron gas by reflecting electron waves emitted from a quantum point contact (QPC) with a circular mirror. Images of electron flow obtained with a scanning probe microscope at liquid He temperatures show interference fringes when the mirror is energized. A quantum phase shifter was created by moving the mirror via its gate voltage, and an interferometric spectrometer can be formed by sweeping the tip over many wavelengths. Experiments and theory demonstrate that the interference signal is robust against thermal averaging.

Address correspondence to R. M. Westervelt

Email: westervelt@deas.harvard.edu.

date: February 23, 2005

Coherent electronics that rely on the phase of electron waves in addition to their amplitude provide new opportunities for sensing and computation, in fields ranging from spintronics to quantum information processing. New techniques based on the interference of electron waves are needed to sense the behavior of electron waves on suitably short length and time scales. Scanning probe microscopy (SPM) has been used to image electron flow in a two-dimensional electron gas (2DEG) in a semiconductor heterostructure without an applied magnetic field [1-9], and in the quantum hall regime [10-15]. At liquid He temperatures, SPM images of the coherent flow of electron waves showing interference fringes have been obtained [1-7].

Electron waves can travel through a two-dimensional electron gas at low temperatures for many microns before losing track of their initial momentum. Their coherence is maintained over distances up to the phase coherence length $\ell_\phi$ [16]. This coherent flow allows one to create an electron interferometer. At finite temperatures the thermal average over the energies of different electrons tends to wash out interference fringes - electrons with different wavelengths produce fringes with different spacings. As a result, averaging can blur out fringes in images bigger than the ballistic thermal length $\ell_T = \hbar v_F / \pi k_B T$, where $v_F$ is the Fermi velocity and $T$ is the temperature. For the measurements reported below at $T$ = 4.2 K, the thermal length is $\ell_T \cong 170$ nm. Interference can still occur for longer electron paths, because $\ell_\phi$ is longer than $\ell_T$, and the coherence of individual electrons is maintained. Collisions with other particles are required to destroy coherence, through electron-electron collisions in our case.

In this letter, we demonstrate the operation of an imaging electron interferometer at liquid helium temperatures. One leg of the interferometer is composed of a quantum point contact (QPC) and a circular electron mirror created by a reflector gate. The other leg is created by the

QPC and a charged SPM tip at a different location than the mirror. Energizing the mirror creates strong interference fringes in the SPM images. Moving the mirror by changing its gate voltage creates a quantum phase shifter. Interferometric spectrometry can be done by moving the SPM tip over many wavelengths. We show experimentally and theoretically that the interference signal is robust against thermal averaging - coherence is still present at relatively high temperatures where $\ell_T$ is smaller than the size of the interferometer. Thermal averaging allows us to distinguish between one-bounce and two-bounce QPC-to-mirror trajectories. The ability to sense and control the phase of electron waves makes interferometric techniques promising for coherent electronics.

Figure 1(a) illustrates the technique we use to image electron flow in a 2DEG at liquid helium temperatures using a SPM [1-7]. The figure shows a SPM tip held above a GaAs/AlGaAs heterostructure that contains a 2DEG below the surface; a QPC is formed by two gates. To image electron flow from the QPC, a negative voltage is applied between the SPM tip and the 2DEG. A small, depleted 'divot' is created immediately below the tip that scatters electron waves arriving from the QPC. Waves backscattered by the divot to the QPC along the same path by which they arrived, pass back through the QPC and reduce its conductance by an amount *ΔG* [1-3]. Waves scattered sideways stay on the same side of the QPC and leave its conductance unchanged. In this way, an image of electron flow can be obtained by recording *ΔG* as the tip is scanned above the sample. The spatial resolution is high, because a single well-defined path between the QPC and the divot beneath the tip determines *ΔG* [1-3,6].

An imaging electron interferometer was constructed by adding a gate to reflect electron waves back to the QPC. Scanning electron micrographs of the sample in Figs. 1(b) and 1(c) show the QPC and a circular reflector gate 1 μm away. An electron mirror is created when a negative voltage between the gate and the 2DEG fully depletes the electron gas below. The interferometer was made using a GaAs/AlGaAs heterostructure grown by molecular beam epitaxy on an n–type GaAs substrate with the following layers: smoothing superlattice, 1 μm GaAs, 22 nm $Al_{0.3}Ga_{0.7}As$, δ-layer of Si donors, 30 nm $Al_{0.3}Ga_{0.7}As$ and a 5 nm GaAs cap. A 2DEG is located 57 nm below the surface. At liquid helium temperatures, the density is $4.2 \times 10^{11}$ $cm^{-2}$, the Fermi energy $E_F = 15$ meV, and the mobility is $1.0 \times 10^6$ $cm^2/Vsec$. All of the SPM images in this paper were recorded at $T = 4.2$ K.

An imaging interferometer for electron waves with a V-shaped trajectory is shown in Fig. 1(b). Each electron leaving the QPC simultaneously travels along two paths - the roundtrip between the QPC and the divot under the SPM tip - the upper leg of the V - and the roundtrip between the QPC and the circular mirror - the lower leg. The backscattered electron waves returning to the QPC along both legs interfere at the QPC, producing strong interference fringes in the SPM image. A trajectory with a double-bounce roundtrip between the QPC and mirror is shown in Fig. 1(c). The high spatial resolution of the SPM and the short transit times for electrons traveling through the device allow interferometric measurements of electron waves at very high frequencies $E_F/h \sim 3$THz with very small time differences $\hbar/E_F \sim 50$fsec, values that are difficult to achieve using conventional methods.

The SPM images in Fig. 2(a-f) show how the circular electron mirror formed by the reflector gate creates an imaging electron interferometer. The series of images in Figs. 2(a-c) were recorded in the green box of Fig. 1(b), at about the same distance (1 μm) from the QPC as the reflector gate, for increasingly negative reflector gate voltages $V_r$: (a) 0.0 V, (b) – 0.4 V and (c) – 0.8 V. The interference fringes increase strongly in amplitude as the electron mirror is formed by depleting the 2DEG below the reflector gate. This demonstrates that the interference occurs between the two green paths in Fig. 1(b): the roundtrip between the QPC and the mirror, and the roundtrip between the QPC and the tip. The images in Figs. 2(d–f) were recorded in the blue box of Fig. 1(c), at about twice the distance (2 µm) from the QPC as the reflector gate, for voltages $V_r$: (d) 0.0 V, (e) – 0.4 V and (f) – 0.8 V. Again, energizing the reflector gate to create the interferometer strongly increases the amplitude of the fringes. As shown below, the fringes in Figs. 2(d-f) were created by interference between the two blue paths in Fig. 1(c): a double-bounce roundtrip between the QPC and the mirror, and a single-bounce roundtrip between the QPC and the tip.

Figure 3 shows how the interferometer can be used as a quantum phase shifter for electron waves; the images also show that the two legs of the interferometer must have comparable lengths to produce a signal at liquid He temperatures. Two series of electron flow images were recorded as the mirror position was moved over a fixed distance by changing $V_r$ from -0.72 V to -0.80 V in steps of -0.02 V. The images in Figs. 3(a-e) were recorded in an area at about the same distance (1 µm) from the QPC as the reflector gate, while the images in Figs. 3(f-j) were recorded at twice that distance. The fringes in both series of images are spaced by half the Fermi wavelength, and they move together as a group. In Figs. 3(a-e), the interference fringes move the same distance as the mirror at an average rate 100 nm/V. But in Figs. 3(f-j), the fringes move twice as far as the mirror at an average rate 230 nm/V. Theoretical simulations presented below show this factor-of-two speedup occurs because the interferometer signal is composed of interfering electron waves that survive thermal averaging. When the SPM tip is twice as far away from the QPC as the mirror, the interfering electron waves must make a double-bounce roundtrip between the QPC and the mirror, to travel the same distance as a single-bounce roundtrip between the QPC and SPM tip, as shown in Fig. 1(c). The interferometer can distinguish between single-bounce and double-bounce roundtrips only when the thermal length $\ell_T$ is shorter than the roundtrip distance between the QPC and the tip, as discussed below.

We can use a remarkable and apparently unnoticed equivalence between thermal averaging and coherent wavepacket dynamics to interpret these results. Suppose we inject a spatially localized, coherent wavepacket through the QPC, which has one open transverse mode. We take the energy profile to match that of the derivative of the Fermi function $-\partial f(E)/\partial E \sim \text{Sech}^2[(E - E_F)/k_B T]$. The wavepacket in the QPC along the longitudinal axis x is then $\psi_T(x) = \int \text{Sech}[(E - E_F)/k_B T] e^{ikx} dk$ where $E = \hbar^2 k^2/2m_e$. In the transverse direction y, it is taken in the transverse eigenmode. The energy profile of the resulting wavepacket is approximately Gaussian, with a spatial uncertainty equal to the thermal length $\ell_T$. It can be shown that the fraction of this wavepacket that eventually succeeds in being transmitted through the QPC is proportional to the thermal conductance [17]. The basic idea is that the relative phase of different energy eigenfuctions is washed out upon time averaging. Strict degeneracy, which would not be removed by time averaging, is prevented by requiring that only one transverse mode be active. Thus understanding the conductance reduces to understanding what restricts or enhances wavepacket backflow through the QPC in the time domain. The key is to notice that wavepackets returning to the QPC along distinct paths, but arriving at different times, cannot interfere with each other; in a quasi-ballistic system, this means the wavepackets must return after a journey of the same length, to within the ballistic thermal length $\ell_T$. It is this interference which gives rise to the fringing as backscattering objects such as the SPM tip are moved, changing the return time and phase of their reflected amplitude.

This provides a time-domain explanation for the persistence of the interference fringes seen in earlier images of electron flow [1-6]. Impurities that backscatter directly to the QPC must reside at the same distance from the QPC as the tip, within an annulus of width $\ell_T$, in order to contribute to the interference. Stronger scatterers such as a concave electron mirror, however, can reflect considerable amplitude back to the QPC, from which it can bounce back to be reflected again: strong fringes are observed in the interferometer for both single- and double-bounce roundtrips between QPC and the mirror. A double-bounce QPC-to-mirror roundtrip can only interfere with a single-bounce QPC-to-divot roundtrip if the SPM tip is approximately twice as far away as the mirror (Fig. 1(c)), so that the two roundtrip paths have the same length. It is easy to see that the interference fringes for a double-bounce QPC-to-mirror roundtrip (Figs. 3(f–j)) should move twice as fast with mirror position. The absence of fringes moving at the same speed as the mirror in Fig. 3(f-j) shows the interferometer is imaging only the double-bounce paths.

The imaging interferometer operates according to the simple model described above in this experiment, because the spatial uncertainly of a thermal wavepacket set by the thermal length $\ell_T$ is much shorter than the interfering electron paths, which are 2 µm long, or longer. This simple picture becomes more complicated at very low temperatures $T < 0.35$ K where $\ell_T$ is longer than the path length. In this case, interference from many different paths for bouncing electrons can contribute to the images, including unwanted multi-bounce paths that hit objects other than the QPC and the mirror, and the interpretation is no longer straightforward. Fringes from the single- and double-bounce QPC-to-mirror roundtrips shown in Figs. 2 and 3 would no longer be separately visible, and interference patterns with different shapes would appear. It might be possible to separately identify single- and double-bounce fringes at very low temperatures by acquiring a series of images for different mirror positions, and then performing image analysis to track how the fringes move. This procedure is not necessary at higher $T$ where

$\ell_T$ is shorter than the path length. The spatial and temporal resolutions determined by $\ell_T$ increase with temperature; an upper limit occurs when $\ell_\phi$ becomes shorter than the path length.

Figure 4 shows quantum-mechanical simulations of the electron flow demonstrating the operation of the interferometer. The transmission through the QPC is calculated using an inverse Greens function technique [8] and a two-dimensional potential that simulates the experimental system [3]. Each pixel in the image is found by calculating the thermally averaged conductance with the tip at that position. Figures 4(a) and 4(b) are at the same distance from the QPC as the reflector gate, while Figs. 4(c) and 4(d) are at twice that distance. The simulation in Fig. 4(a) without the resonator gate energized shows only weak interference fringes caused by backscattering from charged impurity atoms. When the reflector gate is energized in Fig. 4(b), strong interference fringes are formed by waves backscattered from the mirror and from the tip. When the tip is located at twice the distance from the QPC as the reflector gate in Figs. 4(c) and 4(d), similar behavior occurs.

In conclusion, an imaging electron interferometer was constructed in a two-dimensional electron gas - electron waves traveling from a QPC are backscattered by a circular mirror and by the depleted divot beneath a SPM tip, and interfere when they return to the QPC. Strong fringes are produced in images of electron flow when the electron paths have commensurate lengths, within the ballistic thermal length $\ell_T$. Warmer experiments above 4.2 K are certainly desirable. Of course, incoherent electron-electron scattering [17, 18] will eventually become dominant. Before that, however, even better time and spatial resolution will emerge.

This work was supported at Harvard by ARO grant W911NF-04-1-0343, ONR grant N00014-95-1-0104, the Nanoscale Science and Engineering Center (NSF grant PHY-0117795), NSF grant CHE-0073544; and at UCSB by the NSF Science and Technology Center QUEST.

**Figure Captions**

Fig. 1 (a) Schematic diagram showing the technique used to image electron flow through a 2DEG. A negatively biased tip causes backscattering through the QPC. The conductance as a function of tip position is measured to produce an image of electron flow. (b and c) Scanning electron microscope images of the device used to probe the interference fringes, with colored areas indicating where images of electron flow are acquired. The QPC and reflector gate are shown in yellow. The arrows represent the paths that contribute to the interference fringes seen in the images.

Fig. 2 (a-c) Images of electron flow taken in the green area of Fig. 1(b) at the distance of the reflector gate from the QPC for three reflector gate voltages $V_r$: (a) 0.0 V, (b) -0.4 V and (c) – 0.8V showing that the interference fringes are strongly enhanced by energizing the reflector gate. (d-f) Images taken in the blue area of Fig. 1(c) at twice the distance of the reflector gate for voltages $V_r$: (d) 0.0 V, (e) -0.4 V and (f) -0.8 V, showing a similar enhancement. The scale bars are 50 nm long.

Fig. 3 Two series of images of electron flow showing how the interference fringes move as the reflector position is moved a fixed distance by changing $V_r$ from -0.72 V to -0.80 V in steps of – 0.02 V: (a-e) images recorded at the same distance as the reflector gate; (f-j) images recorded at approximately twice that distance. In (a–e) the interference fringes move the same distance as the reflector – the center dot moves from a peak to a valley, but in (f-j) the fringes move twice as far as the reflector – the center dot moves from a peak to a valley, then to the next peak. The scale bars are 50 nm long.

Fig. 4 Quantum mechanical simulations of the electron flow through the device: (a) and (b) are taken at the same distance from the QPC as the reflector gate; (c) and (d) are at twice the distance of the reflector gate. Before the interferometer is formed, images (a) and (c), only weak interference fringes occur. When electron waves are reflected from the mirror, (b) and (d), strong interference fringes are created. The scale bars are 50 nm long.

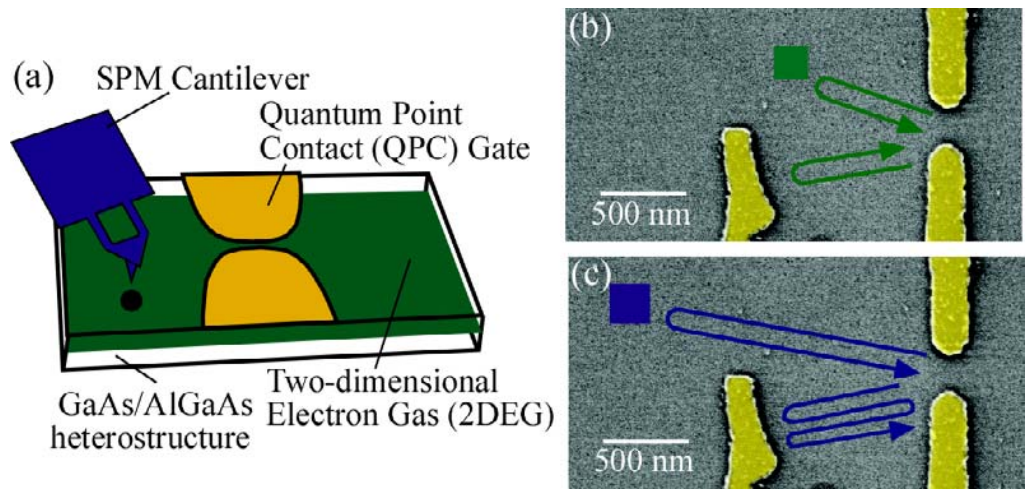

Figure 1

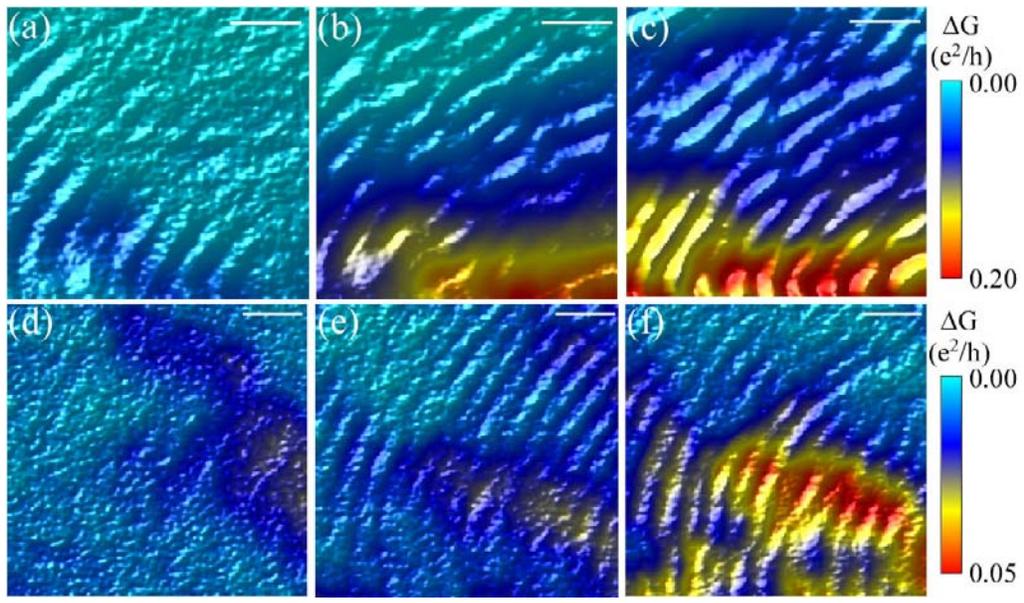

Figure 2

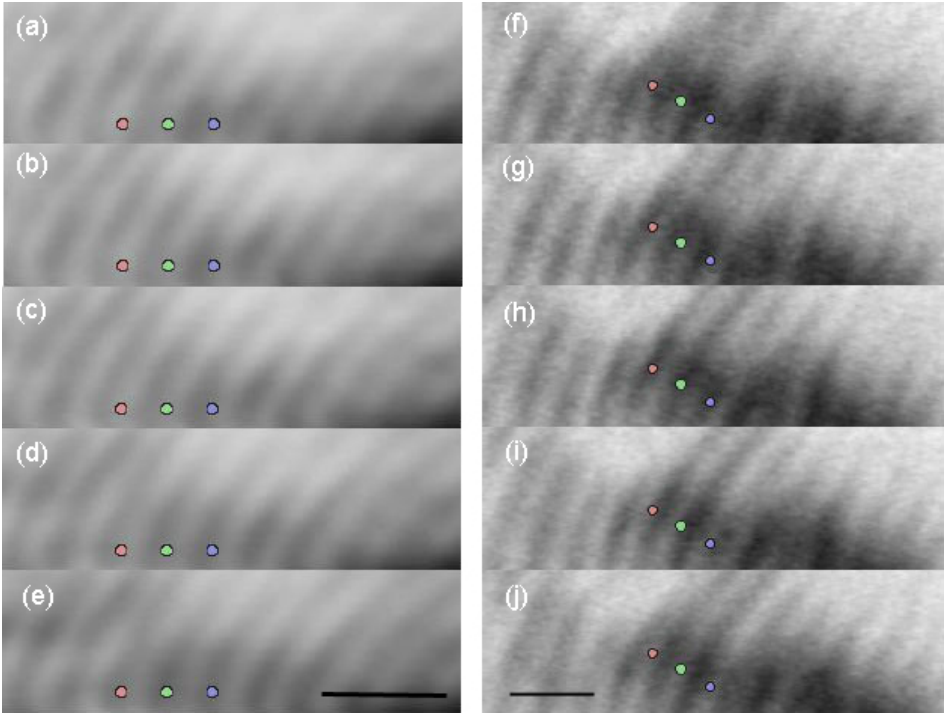

Figure 3

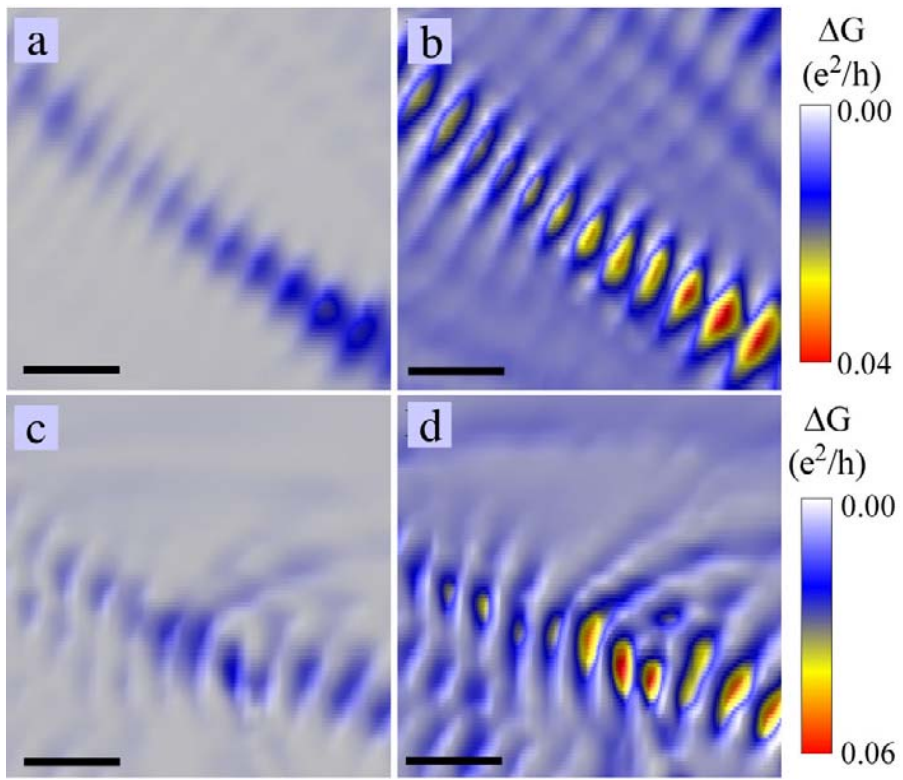

Figure 4